\documentclass[11pt]{article}

\usepackage{amsmath,amssymb}
\usepackage{here}
\usepackage{graphicx,epsfig}
\usepackage{citesort}
\usepackage{psfrag}

\newcommand{\be}{\begin{eqnarray*}}
\newcommand{\ee}{\end{eqnarray*}}
\newcommand{\bed}{\begin{description}}
\newcommand{\eed}{\end{description}}
\newcommand{\bea}{\begin{eqnarray}}
\newcommand{\eea}{\end{eqnarray}}
\newcommand{\beq}{\begin{equation}}
\newcommand{\eeq}{\end{equation}}

\begin{document} 

\hfill{\small FZJ--IKP(TH)--2007--12} \\[1.8em]
\begin{center}
\Large{\bf Pion reactions on two--nucleon systems}

\vspace{5mm}
\large{C. Hanhart}

\vspace{5mm}
\footnotesize{Institut f\"{u}r Kernphysik, \\
Forschungszentrum J\"{u}lich GmbH, D--52425 J\"{u}lich, Germany}

\vspace{5mm}
\begin{abstract}
  We review recent progress in our understanding of elastic and inelastic pion
  reactions on two nucleon systems from the point of view of effective field
  theory. The discussion includes $\pi d$ scattering, $\gamma d\to \pi^+ nn$,
  and $NN\to NN\pi$. At the end some remarks are made on strangeness production
reactions like
  $\gamma d\to K\Lambda N$ and $NN\to K\Lambda N$.
\end{abstract}

\end{center}

\section{Introduction}

Even after several decades of research, 
the interactions and dynamics of strongly
interacting few--nucleon systems are still not
fully understood. Although phenomenological
approaches are quite often very successful
in describing certain sets of data, a coherent
overall picture with clear connection
to the fundamental theory, QCD, is still
lacking. 

The only way to a systematic and well controlled understanding of hadron
physics is through the use of an effective field theory. The effective field
theory (EFT) for the Standard Model at low energies is chiral perturbation
theory (ChPT).  This EFT already provided deep insights into strong
interaction physics at low energies from systematic studys of the
$\pi\pi$~\cite{gilberto} and the $\pi N$~\cite{ulfs,nadiaq3} system ---
reviewed in Ref.~\cite{bastian} --- as well as the $NN$
system~\cite{birapaulo,evgenirev,machleidt}.  Here we focus on the field of
elastic and inelastic reactions on few nucleon system.

A first step towards a systematic study of elastic and inelastic reactions
on nuclei was taken by Weinberg already in 1992 \cite{wein}. He suggested that
all that needs to be done is to convolute transition operators, calculated
perturbatively in standard ChPT, with proper
nuclear wave functions to account for the non--perturbative character of the
few--nucleon systems. This procedure looks very similar to the so--called
distorted wave born approximation used routinely in phenomenological
calculations, but, in contrast to this, opens up the possibility
to use a power counting scheme.
  Within ChPT this idea was already applied to a large number of
reactions like $\pi d\to \pi d$ \cite{beane}, $\gamma d\to \pi^0 d$
\cite{kbl,krebs}, $\pi {^3}$He$\to \pi {^3}$He \cite{baru}, $\pi^- d\to \gamma
nn$ \cite{garde}, and $\gamma d\to \pi^+ nn$ \cite{lensky}, where only the
most recent references were given.
We start our presentation with a brief description of $\pi d$ scattering
in sec.~\ref{pidintro}.

Using standard ChPT especially means to use an expansion
in inverse powers of $M_N$ --- the nucleon mass. However,
some pion--few-nucleon diagrams employ few--body singularities
that lead to contributions non--analytic in $m_\pi/M_N$, with
$m_\pi$ for the pion mass~\cite{recoils}. This is discussed
in detail in sec.~\ref{cuts}. There we show that the appearance
of these contributions is linked to the Pauli principle operative
while there is a pion in flight. In this context we discuss 
also the reaction $\gamma d\to \pi^+ nn$, for in the mentioned
sense this reaction is complementary to $\pi d$ scattering.
We show that within ChPT the existing data can be described 
to high accuracy
and therefore the reaction qualifies as a tool to measure the
$nn$ scattering length.

A problem was observed when the original scheme by Weinberg was
applied to the reactions $NN\to NN\pi$ \cite{park,unserd}: the
inclusion of potentially higher order corrections were large and lead
to even larger disagreement between theory and experiment than found
in earlier phenomenological studies~\cite{haidenbauermachner}.  For the
reaction $pp\to pp\pi^0$ one loop diagrams that in the Weinberg
counting appear only at NNLO where evaluated \cite{dmit,ando} and they
turned out to give even larger corrections putting into question the
convergence of the whole series.  However, already quite early the
authors of Refs. \cite{bira1,rocha} stressed that an additional new
scale enters for $NN\to
NN\pi$ that needs to be accounted for in the power counting.
 Since the two nucleons in the initial state need to have
sufficiently high kinetic energy to put the pion in the final state
on--shell, the initial momentum needs to be larger than
\begin{equation}
p_{thr} = \sqrt{M_Nm_\pi} \ .
\label{pthr}
\end{equation}
The proper way to include this scale was presented in Ref. \cite{ch3body} ---
for a recent review see Ref.  \cite{report}. As a result, pion $p$-waves are
given by tree level diagrams up to NLO and the corresponding calculations
showed satisfying agreement with the data. However, for pion $s$--waves loops
appear already at NLO~\cite{ch3body,withnorbert}.  In sec.~\ref{nnpi} we
discuss their effect on the reaction $NN\to d\pi$ near threshold. In some
detail we will compare the effective field theory result to that on
phenomenological calculations.

The central concept to be used in the construction of the
transition operators is that of reducibility, for it allows
one to disentangle effects of the wave functions and those from
the transition operators. As long as the operators are energy
independent, the scheme can be applied straight
forwardly~\cite{wallacephillips},
however, as we will see below, for energy dependent interactions
more care is necessary. This will also be subject of sec.~\ref{nnpi}.

Once the reaction $NN\to d\pi$ is understood within effective field theory
one is in the position to also calculate the so--called dispersive and
absorptive corrections to the $\pi d$ scattering length. This
calculation will be presented in section~\ref{pid}. 

When switching to systems with strangeness one immediately observes that the
initial momentum needs to be quite large in any production reaction where
there are two nucleons and no strangeness in the initial state. Strangeness
conservation demands that at least two particles with strangeness in the final
state. Therefore the mass produced is at least
$$
\Delta=M_\Lambda-M_N+m_K-m_i \ ,
$$
where $m_i$ denotes the mass present in the initial state in addition to
the two nucleons, e.g., $m_i\simeq m_\pi$ for $\pi d\to K\Lambda N$ and
$m_i=0$ for $NN\to N\Lambda K$. Therefore, in the latter reaction
the expansion parameter of the ChPT, namely
 the
initial momentum in units of the nucleon mass, is 0.85, using the analog
of Eq.~(\ref{pthr}). Also in the former reaction we are faced with an initial
momentum 0.6 in units of the nucleon mass --- again not useful as an expansion parameter.
 Clearly, in those cases the chiral expansion as proposed above is no longer
applicable. In sec.~\ref{YN} we discuss, how for those large momentum transfer
reactions the scattering lengths of the outgoing baryons can still be 
extracted in a controlled way using dispersion theory.

We close with a brief summary and outlook.

\section{Remarks on the $\pi d$ system}
\label{pidintro}

\begin{figure}[t!]
\begin{center}
\epsfig{file=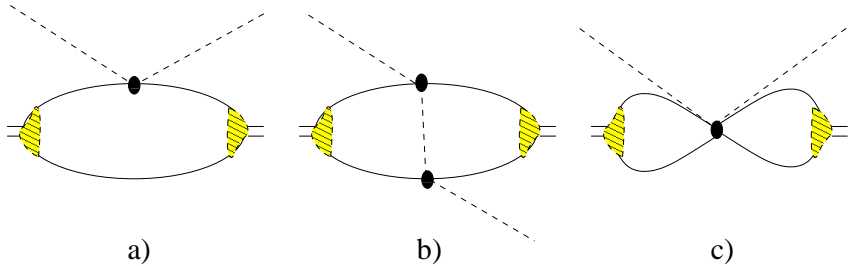,height=3.5cm}
\end{center}
\caption{Typical diagrams that contribute to $\pi d$ scattering.
Shown is the one--body term (diagram $(a)$), the leading
two--body correction (diagram $(b)$) and a
possible short--ranged operator (diagram $(c)$).}
\label{leadingpid}
\end{figure}

We start our discussion with some remarks on the $\pi d$ system.
Pion-deuteron ($\pi d$) scattering near threshold plays an exceptional role in
the quest for the isoscalar $\pi N$ scattering length $a_+$, since the
deuteron is an isoscalar target. Therefore one may write $\mbox{Re}(a_{\pi d})
= 2a_+ + (\mbox{few--body corrections}) \ .$ The first term $\sim a_+$ is
simply generated from the impulse approximation (scattering off the proton and
off the neutron; diagram $(a)$ of Fig.~\ref{leadingpid}) and is independent of
the deuteron structure.  Thus, if one is able to calculate the few--body
corrections in a controlled way, $\pi d$ scattering is a prime reaction to
extract $a_+$ (most effectively in combination with an analysis of the high
accuracy data on pionic hydrogen).  In addition, already at threshold the $\pi
d$ scattering length is a complex-valued quantity. It is therefore also
important to gain a precise understanding of its imaginary part - this issue
will be discussed in sec.~\ref{pid}.

Recently the $\pi d$ scattering length was measured to be \cite{PSI1}
\be
a_{\pi d}^{\mbox{exp}} =\left (-26.1\pm 0.5\mbox \, + \, i (6.3\pm 0.7)\right )\times
\,10^{-3} \ m_\pi^{-1} \ , 
\label{exp}
\ee 
where $m_\pi$ denotes the mass of the charged
pion. In the near future a new measurement with a projected total uncertainty of 0.5\% for
the real part and 4\% for the imaginary part of the scattering length will be
performed at PSI \cite{detlev}. Clearly, performing calculations up to this accuracy 
poses a challenge to theory that several groups recently took up
\cite{BBEMP,rus,doeringoset,arriola,mitandreas,danielneu}. In addition, an
interesting isospin violating effect in pionic deuterium was found, see \cite{MRR}.
 For a review on older
work we refer to Ref. \cite{al}.

A typical few--body correction to the $\pi d$ scattering length is shown in
diagram $(b)$ of Fig.~\ref{leadingpid}. As we will see below, the contribution
of this diagram largely exhausts the value of the $\pi d$ scattering length
not leaving much room for a contribution from $a_+$, or stated differently,
pointing at a small value of $a_+$.  Based on calculations within pion less
EFT, it was claimed recently that this diagram is sensitive to the short range
part of the deuteron wave function~\cite{gries,silasmartin}. As a consequence
field theoretic consistency requires that at the same order there is to be a
local operator to absorb this model dependence --- the corresponding diagram
is shown as diagram $(c)$. Since this diagram comes with an a priori unknown
strength not fixed by symmetries, $\pi d$ scattering would be useless for the
extraction of $a_+$. However, systematic investigations showed that, as soon
as the pion exchange is included explicitly in the $NN$ potential, diagram
$(b)$ can be evaluated in a controlled
way~\cite{rus,arriola,mitandreas,danielneu}. Given this we assume from now on
that short--ranged operators contributing to the $\pi NN\to\pi NN$ transition
potential scale naturally. In other words, contribute with strength parameters
of the order of one. Based on this we may estimate the contribution of diagram
$(c)$ of Fig.~\ref{leadingpid} relative to diagram $(b)$. This kind of
analysis gives a relative suppression of the order $\mathcal{O}(\chi^2)$,
where $\chi=m_\pi/M_N$ is the standard expansion parameter of ChPT with $M_N$
($m_\pi$) for the nucleon (pion) mass. This, together with the knowledge
that diagram $(a)$ largely exhausts the value of the $\pi d$ scattering
length, allows us to estimate the theoretical limit for the extraction of
$a_+$ from a measurement of the $\pi d$ scattering length. We find
\begin{equation}
\Delta a^{\rm theo} \sim  5\times 10^{-4}  \ m_\pi^{-1} \ .
\end{equation}
To meet this theoretical limit we need to include in the calculation
all contributions to the $\pi d$ scattering length lower than
$\mathcal{O}(\chi^2)$.

\begin{figure}[t!]
\begin{center}
\epsfig{file=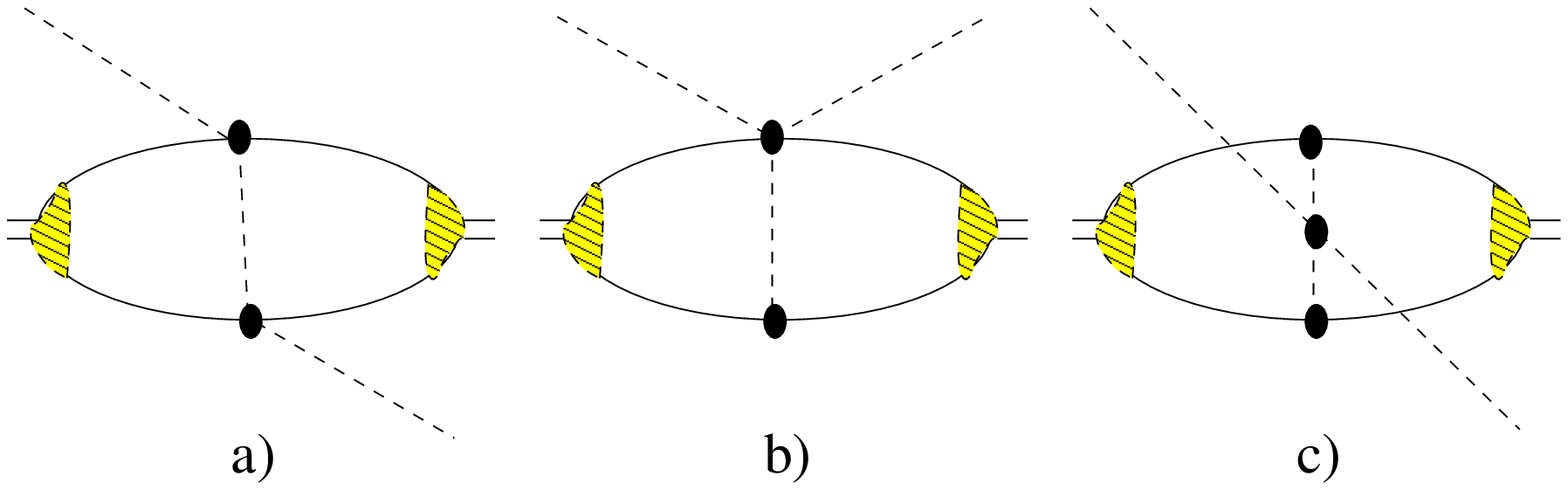,height=3.5cm}
\end{center}
\caption{Formally leading few--body corrections
to the $\pi d$ scattering length.}
\label{fewbodypid}
\end{figure}

Already in his original work, Weinberg discussed $\pi d$ scattering
at threshold as an illustrative example~\cite{wein}. 
As usual, the leading contributions to the transition operators
 are all those tree--level diagrams 
that can be constructed from the leading $\pi N$ and $\pi\pi$
Lagrangians. Those are shown in Fig.~\ref{fewbodypid}.
Note that it is very important that the complete set 
of diagrams is considered, since, for example, both
diagram $(b)$ as well as diagram $(c)$ are depending
on the particular choice made for the pion field.
However, the sum of both is independent of this
choice, as was first pointed out in Ref.~\cite{robilottawilkin}.

Since all diagrams of Fig.~\ref{fewbodypid} contribute to the same chiral
order, naively one expects them to give similar contributions. However,
explicit numerical evaluations showed, that diagram $(a)$ exceeds the sum of
the other two by about two orders of magnitude~\cite{wein}.
Can we understand this? One possible explanation could be that the
sum of $(b)$ and $(c)$ is small because of significant cancellations
between the two, probably due to the mechanism indicated at the end
of the previous paragraph. Another possible explanation was given
in Ref.~\cite{beane}: as a consequence of the small binding energy of
the deuteron, $\epsilon$, the typical nucleon momentum inside the
deuteron, $\gamma$, is also small, $\gamma\sim \sqrt{M_N\epsilon}$,
which turns out to be numerically about 1/3 of the pion mass. Since
diagram $(a)$ is proportional to the expectation value of $1/\vec q\, ^2$,
where $\vec q$ denotes the momentum transfer through the pion, and
the sum of diagram $(b)$ and $(c)$ is proportional to the 
expectation value of $\vec q\, ^2/(\vec q\, ^2+m_\pi^2)^2$, we
expect the ratio of the two contributions to be of the order of
$(\gamma/m_\pi)^4\sim 10^{-2}$, where $\left|\vec q\, \right|$
 was identified 
with the value of $\gamma$ defined above. Thus, the small binding
energy of the deuteron seems to provide a natural explanation 
of the relative suppression of diagram $(b)+(c)$ to $(a)$.
However, more systematic studies are necessary.

Another important issue is the role of nucleon recoil contributions.  All
calculations mentioned so far use as starting point the limit of infinitely
heavy nucleons; corrections due to the finite nucleon mass are then included
as a power expansion in $1/M_N$. However, it was already observed in the
70s~\cite{KK}, based on calculations using Gaussian wave--functions, that this
way one may miss important terms. This was further investigated in
Ref.~\cite{recoils}, where the analysis was done model independently and the
appearance of these additional contributions was related to the Pauli
principle in the intermediate $NN$ state, while the pion is in flight. This
will be discussed in detail in the next section.

\section{Role of $\pi NN$ cuts}
\label{cuts}

\begin{figure}[t]
\begin{center}
\psfrag{xx}{{$\vec P_R$}}
\psfrag{yy}{{$\vec K_\pi$}}
\epsfig{file=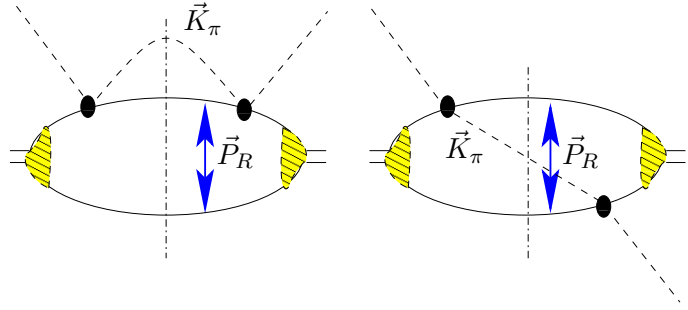, height=4cm, angle=0}
\end{center}
\caption{Typical pion loop contributions to $\pi d$ scattering.
Since the exchange of the two nucleons in the intermediate state
(at the perpendicular lines) transforms one diagram into the other,
the imaginary parts (and their analytic continuation) are
linked by the Pauli principle, as described in the text. }
\label{pinncuts}
\end{figure}

Let us investigate the diagrams of Fig.~\ref{pinncuts}
from the point of view of their $\pi NN$ cut. Then we
may write for the corresponding matrix elements
\begin{equation}
 I_{\pi NN}(Q)=\int \frac{dK_\pi K_\pi^2 \, dP_R P_R^2}{(2\pi)^6}\frac{f(
  K_\pi^2,P_R^2)}{Q-K_\pi^2/(2m_\pi)-P_R^2/M_N + i0} \ ,
\label{full}
\end{equation}
where the function $f$ contains the reaction specific parts like vertex
functions, wave functions and transition operators. The only part
spelled out explicitly is the $\pi NN$ propagator, where for 
simplicity non--relativistic kinematics was chosen. Here
$Q$ denotes the excess energy with respect to the $\pi NN$
threshold. The goal of this study is to compare the full expression
given in Eq.~\ref{full} with the corresponding one that
emerges when static nucleons are used. This corresponds to taking
the limit $M_N\to\infty$ prior to integration and we may write
\begin{equation}
 I_{\pi NN}^{(static)}(Q)=\int \frac{dK_\pi K_\pi^2 \, dP_R P_R^2}{(2\pi)^6}\frac{f(
  K_\pi^2,P_R^2)}{Q-K_\pi^2/(2m_\pi) + i0} + O\left(\frac{m_\pi}{M_N}\right) \ .
\label{full_static}
\end{equation}

It was generally assumed that the difference between the integrals of
Eqs.~(\ref{full}) and (\ref{full_static}) can be accounted for in a
polynomial expansion in $m_\pi/M_N$. However, this is in general not
the case: at pion production threshold $I_{\pi NN}(Q)$ has a branch
point singularity. As we will see this leads to a contribution
non--analytic in $m_\pi/M_N$, even below pion threshold. To see this,
let us study in detail the cut contribution by replacing the $\pi NN$
propagator by its delta function piece
$$
(Q-K_\pi^2/(2m_\pi)-P_R^2/M_N+i0)^{-1} \ \to \ -i\pi \delta
(Q-K_\pi^2/(2m_\pi)-P_R^2/M_N) \ .
$$
Then we may write
$$
 I_{\pi NN}^{(cut)}(Q) = -i\pi m_\pi \int
\frac{dP_R P_R^2}{(2\pi)^6}f\left(K_\pi^{on}(Q,P_R^2)^2,
P_R^2\right)K_\pi^{on}(Q,P_R^2) \ .
$$
for the full contribution, where
$K_\pi^{on}(Q,P_R^2)=\sqrt{2m_\pi(Q-P_R^2/M_N)}$ and
$$
I_{\pi NN}^{(cut,static)}(Q) = -i\pi m_\pi K_\pi^{on}(Q)
\int \frac{dP_R P_R^2}{(2\pi)^6}f(K_\pi^{on}(Q)^2,
  P_R^2) \ 
$$
for the static one, with $\left.K_\pi^{on}(Q)\right|_{static}=\sqrt{2m_\pi Q}$ .
At this point we see already one important difference between
the static and the full treatment: while the imaginary part
of the former scales in a completely wrong way, namely
according to two--body phase--space, the latter shows
the proper scaling as three--body phase space. 

But this is not all. Also below the $\pi NN$ threshold 
the static contribution gives wrong results. To see
this let us focus on the contribution at 
$\pi NN$ threshold , $Q=0$. Although for this 
kinematics both the above integrals are real,
the presence of the $\pi NN$ cut still plays 
a significant role. To evaluate the relevant integral
the on--shell momentum $K_\pi^{on}(Q,P_R^2)$ needs
to be continued analytically to imaginary values
using the prescription
$$K_\pi^{on}(0,P_R^2)=\sqrt{-2m_\pi P_R^2/M_N}\to i\sqrt{2m_\pi P_R^2/M_N} \ . $$ 
With this we get
$$
 I_{\pi NN}^{(cut)}(0) = \pi m_\pi \sqrt{\frac{2m_\pi}{M_N}} \int
\frac{dP_R P_R^3}{(2\pi)^6}f\left(K_\pi^{on}(0,P_R^2)^2,
P_R^2\right) \ .
$$
whereas
$$
I_{\pi NN}^{(cut,static)}(0) = 0 \ .
$$
Thus, taking the $M_N\to \infty$ limit prior to integration 
is in general not allowed in the presence of few--body cuts.

The natural question is: when does this matter? As explained
the mentioned effect originates from the opening of the
physical $\pi NN$ threshold. However, this threshold can
only matter, if the $\pi NN$ state is allowed by selection rules.
In the isospin limit the two nucleons are identical particles
that are to obey the Pauli principle. It therefore depends on
the operator that acts on the deuteron wave function to produce
the intermediate pion, whether or not the $NN$ state is allowed and,
consequently, whether the above contributions matter.

Let us first look at $\pi d$ scattering. The leading operator
that contributes to the $\pi N\to \pi N$ transition is
the so--called Weinberg--Tomozawa term $\propto \epsilon^{abc}\tau^c$, which is a
vector in isospin space, but spin and momentum independent.
Therefore, this operator acting on the deuteron (isospin 0 and spin 1), leads
to an $NN$ state that is isospin 1 and spin 1, predominantly in an $s$--wave
due to the momentum independence of the transition operator. This
$NN$ state is forbidden by the Pauli principle and therefore all said
in the first part of this section does not matter and the
static approximation gives a good description of the
leading few--body correction.
The same holds, e.g., for $\pi d\to \gamma NN$.

However, there are reactions where we expect the 
above terms to become significant. One example
is the reaction $\gamma d\to\pi^+ nn$ that will be discussed
in more detail in the next section. Here the operator
acting on the deuteron wave function in leading order
 is the so--called
Kroll--Ruderman term, which is a vector in both isospin 
as well as spin space. Thus, a transition to an $NN$ pair
in the $^1S_0$ isovector state, which is allowed by the Pauli 
principle, is possible. This reaction was studied in detail
in Ref.~\cite{recoils2}, and indeed the pattern sketched above
on general grounds was observed. However, for a Pauli allowed
intermediate state, the two nucleons will interact. It was
found that the inclusion of the two--nucleon intermediate state
gives a significant contribution, however, numerically smaller
than the static exchange itself. This is different to the
case of $\pi d$ scattering with an isoscalar $\pi N$ interaction 
that also leads to a Pauli allowed intermediate state. In this
case the inclusion of the $NN$ interaction at threshold numerically
restored the contribution of the static exchange~\cite{faeldt}.

\section{The reaction  $\gamma d\to\pi^+ nn$}

\begin{figure}[t]
\begin{center}
\psfrag{xx}{{$\vec P_R$}}
\psfrag{yy}{{$\vec K_\pi$}}
\epsfig{file=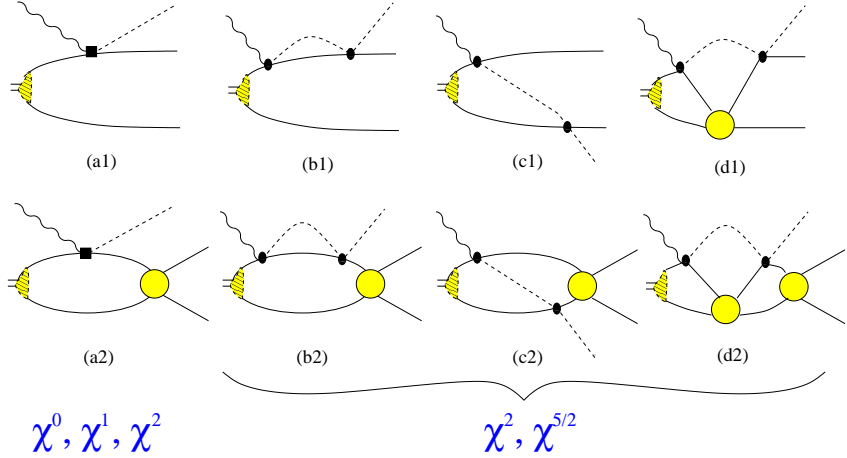, height=6cm, angle=0}
\end{center}
\caption{Diagrams contributing to $\gamma d\to\pi^+nn$ up
to order $\chi^{5/2}$. As before, solid lines denote nucleons,
and dashed lines pions.  The wavy lines denote photons. 
The hatched areas denote the deuteron wave function and the filled
circles the $NN$ interaction.}
\label{gammad}
\end{figure}

The reaction $\gamma d\to\pi^+ nn$ was studied
intensively already in the 70s --- for a review
see Ref.~\cite{laget}. At this time only diagrams
$(a1)$ and $(a2)$ of Fig.~\ref{gammad} were included.
In Ref.~\cite{laget} there is only one comment to
an unpublished work, where pion rescattering (diagrams $(c1)$ and
$(c2)$ of Fig.~\ref{gammad})
was calculated, however, in the static approximation.
It is stated that the inclusion of this contribution
destroys the nice agreement of the calculation
based on the one--body terms only and therefore
it will no longer be considered.
Based on the discussion of the previous section we
now understand, why the static pion exchange diagram
gave a contribution way too large: in a complete
calculation it would have been largely canceled
by the recoil corrections, since the two--nucleon
state in diagrams $(b)$, $(c)$ and $(d)$ can go
on--shell while the pion is in flight.

\begin{figure}[t]
\begin{center}
\epsfig{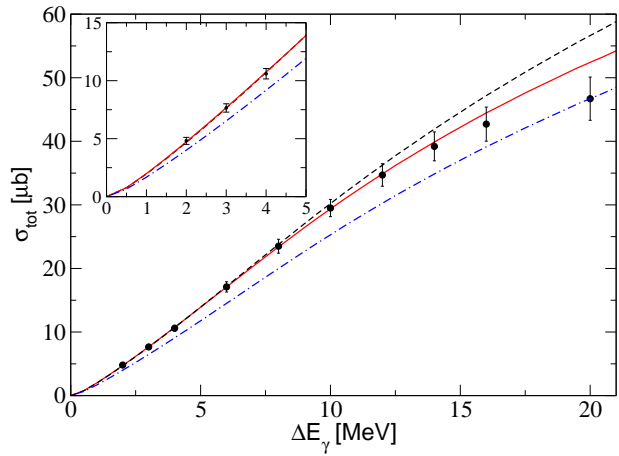}
\caption{ Total cross section of the reaction $\gamma d\to \pi^+ nn$
at LO (dashed line), NLO (dash--dotted line) and $\chi^{5/2}$ order (solid line)
%NNLO (solid line)
 together with experimental data from Ref.~\cite{MIT}.
}
\label{gammad_results}
\end{center}
\end{figure}

The results of our calculation are shown in Fig.~\ref{gammad_results}.  At
leading order and next--to--leading order only one--body terms contribute
(c.f. Fig.~\ref{gammad}) with their strength fixed by the chiral Lagrangian.
The relevant $\gamma p\to\pi^+n$ vertices are momentum independent in both
cases and therefore their energy dependence is identical. At NNLO there is a
counter term for the transition $\gamma p\to\pi^+n$ and the strength of the
one--body operator can be adjusted to data~\cite{threspiprod}. This gives a
large fraction of the shift in strength when going from NLO to NNLO. In
addition the amplitude gets energy dependent~\cite{heavyChPT}. Another source
of energy dependence comes from the few--body corrections as well as higher
partial waves that start to contribute at this order. As can be seen from the
figure, the data is described very nicely in the whole low energy region
considered.

It seems as if the few body corrections, when treated properly, only
have a minor effect on the event rates for $\gamma d\to\pi^+ nn$, however,
this is correct only for the total cross section. Especially the 
neutron momentum distributions are sensitive to higher order corrections
and those are to be understood to very
high accuracy, in order to make use of this reaction to extract
the $nn$ scattering length.

%\begin{itemize}
%\item Nucleon \bl{recoil} corrections can be \textcolor{blue}{numerically
%    significant}; they scale as \re{$\sqrt{m_\pi/M_N}$}
%    \textcolor{green1}{(V.M. Kolyabasov et al. (1972))}
%\item Even nucleons in the intermediate state need to \bl{fullfil the Pauli
%    principle};
%this connects rescattering terms to one--body terms
%\textcolor{green1}{(M.P. Rekalo et al. (2002))}
%
%\re{$\to$ no net few body corrections for $\pi d$}
%\end{itemize}

On the level of neutron momentum distributions diagram $(a1)$ leads
to very specific signals due to the so--called quasi--free production.
When all particles in the final state go forward, the intermediate 
proton is very near on--shell and the diagram gives a large contribution,
which decreases quickly,
however, as we go away from forward kinematics.
In addition, the quasi--free production favors large relative momenta
of the two neutrons. Near threshold this is clear, as --- in the center
of mass system --- the spectator neutron keeps on going with half the
deuteron momentum, whereas the reaction neutron gets decelerated to almost
at rest through the production process.
 
\begin{figure}[t]
\begin{center}
\epsfig{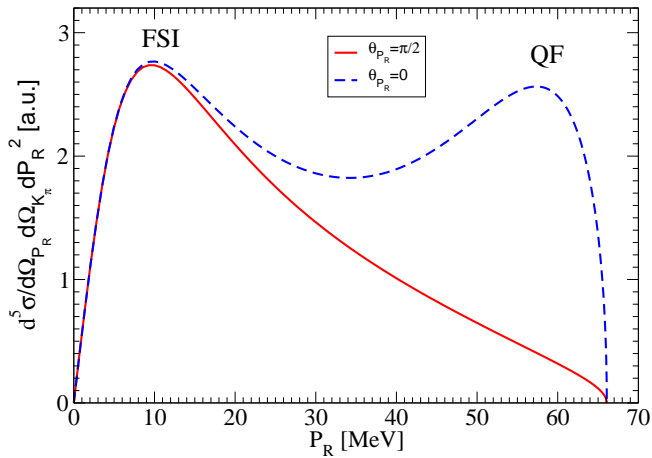}
\caption{Predicted event rate for the reaction $\gamma d\to \pi^+ nn$
at an excitation energy of 5 MeV
as a function of the $nn$ relative momentum $P_R$ for two different orientations
of it.  The region of small values of $P_R$ are dominated by the $nn$ final
state
interaction (FSI) and that of large values of it by the quasi-free production (QF). 
}
\label{gammad_dist}
\end{center}
\end{figure}

All diagrams with a final--state interaction, on the other hand, 
give contributions peaked at small relative $nn$ momenta and almost
insensitive to their orientation.
This is illustrated in Fig.~\ref{gammad_dist}, where the differential
rate is shown for two different angles as a function of the relative
$nn$ momentum $\vec P_R$. The dashed line
denotes the distribution where $\vec P_R$ is directed along the beam
axis, whereas for the solid line it is perpendicular to the beam. In
the former case the distribution shows a clear two--peak structure ---
the quasi--free production shows up at large $P_R$ and the final state
peak shows up at small $P_R$. In the latter case, on the other hand, the
quasi free contribution has disappeared and only the final state interaction
piece remains with basically identical strength.

The hight and shape of the FSI peak is sensitive to the value of
$a_{nn}$---the neutron--neutron scattering length. A systematic
study revealed that high accuracy data on $\gamma d\to \pi^+nn$
will allow one to extract the $nn$ scattering length with an
uncertainty of the order of $0.1$ fm which is compatible with
that estimated for the competing reactions, $pd\to nnp$~\cite{nnfrompd} and
$\pi^- d\to \gamma nn$~\cite{garde}.

\section{$NN\to d\pi$}
\label{nnpi}

As sketched in the introduction, for reactions of the type
$NN\to NN\pi$ a simultaneous expansion in the large
initial momentum $p_{\rm thr}\sim \sqrt{m_\pi M_N}$, that also sets the scale
for the typical momenta in the loops, and the pion
mass is compulsory.
Before we go into details in discussing a particular
reaction channel we would like to briefly illustrate the
impact of this. In practice this means that momenta and pion masses are
to be treated independently in the power counting.

\begin{figure}[t!]
\begin{center}
\epsfig{file=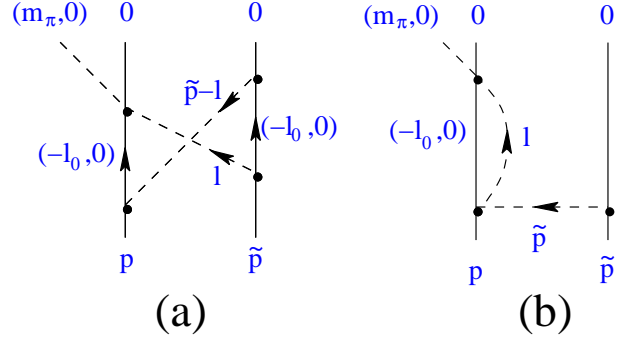,height=4.5cm}
\end{center}
\caption{Two typical pion loops that contribute
to $NN\to NN\pi^+$. Diagram $(a)$ starts to contribute
at NLO whereas diagram $(b)$ starts to contribute at
N$^4$LO.}
\label{power}
\end{figure}

To see how this works let us for example estimate the
contributions of the loops shown in Fig.~\ref{power}.
For diagram $(a)$ we then estimate
$$\frac{p}{f_\pi^2}\left(\frac{p}{f_\pi}\right)^3\left(\frac{1}{p^2}\right)^2
\left(\frac{1}{p}\right)^2\frac{p^4}{(4\pi)^2} \sim {\frac{p^2}{M_N^2}} \, , $$
where the different terms refer to the $\pi N\to \pi N$ vertex, the three
$\pi NN$ vertices, the two pion propagators, the two nucleon propagators, and
the integral measure, in order. Each individual piece was expressed
by the dimensionful parts, where momenta were identified with their typical
values. For more details we refer to Appendix E of Ref.~\cite{report}.
  To come to the order estimate we used $4\pi f_\pi\sim M_N$ and dropped
an overall factor of $1/f_\pi^3$ common to all production amplitudes.
On the other hand we find for diagram $(b)$
$$\left\{\left(\frac{m_\pi}{f_\pi^2}\right)^2
\frac1{m_\pi}\frac{1}{m_\pi^2}\frac{m_\pi^4}{(4\pi)^2}\right\}
\frac1{p^2}\left(\frac{p}{f_\pi}\right)
 \sim {\frac{m_\pi^3}{pM_N^2}} \ . $$
The expression in the curly bracket refers to the pion loop --- it contains
two $\pi N\to \pi N$ vertices, one nucleon and one pion propagator as well as
the integral measure --- however, in this case the typical momentum is of the
order of $m_\pi$ instead of $p_{\rm thr}$ as in the previous example. The
reason
is simply that one may choose in the loop the momenta such that the large
momentum does not run through the pion. Then the large scale does not appear
in the loop at all since the leading heavy baryon propagator feels 
energies only. Outside the curly bracket the momentum transfer is large
and therefore the pion propagator as well as the pion vertex appear with
$p\sim p_{\rm thr}$.

If we compare the two order asignments we observe that under
the assumption that momenta are of order $m_\pi$, both expressions
appear at the same order and are therefore expected to be
of similar order of magnitude. However, if we assume $p$ to be of order
$p_{\rm thr}$, then diagram $(a)$ is of order $m_\pi/M_N$ and 
diagram $(b)$ is of order $(m_\pi/M_N)^{5/2}$, which corresponds
to a relative suppression of $(m_\pi/M_N)^{3/2}\sim 1/20$.  
An explicit calculation~\cite{dmit} revealed an even larger suppression
of diagram $(b)$, which turned out to be suppressed by a factor
of $50$ compared to $(a)$. For more examples we refer to Ref.~\cite{report}.

\begin{figure}[t!]
\begin{center}
\epsfig{file=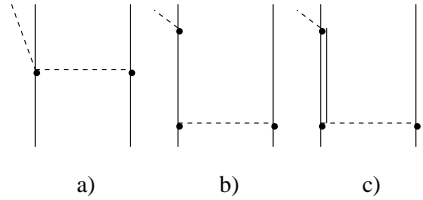,height=2.6cm}
\end{center}
\caption{Tree level diagrams that contribute
to $pp\to d\pi^+$ up to NLO. Solid lines denote 
nucleons, dashed ones pions and the double line 
the propagation of a Delta--isobar.}
\label{tree}
\end{figure}

As already discussed in the context of $\pi d$ scattering,
in some cases only the sum of various diagrams can give
meaningful results, since the individual diagrams depend on
the choice made for the pion field.
Due to the reordering of loop diagrams, the members
of  those invariant subgroups
are not necessarily  of the same chiral order anymore, in contrast
to the original Weinberg counting.
That, regardless this, also the new scheme gives meaningful results 
is demontrated in Ref.~\cite{mitandreasw}.

\begin{figure}[t!]
\begin{center}
\epsfig{file=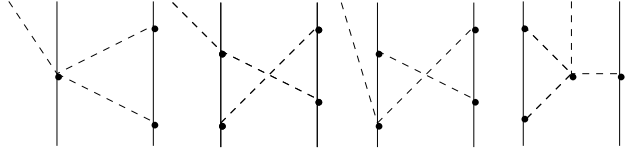,height=2.cm}
\end{center}
\caption{Irreducible pion loops with nucleons
only that start to
contribute to $NN\to NN\pi$ at NLO that were
considered in Ref. \cite{withnorbert}.}
\label{NLOdiags}
\end{figure}

The tree level amplitudes that contribute to $pp\to d\pi^+$ are
shown in Fig. \ref{tree}.
In Ref. \cite{withnorbert} all NLO contributions of loops that start to contribute
to $NN\to NN\pi$ at NLO were\footnote{In a scheme with two expansion parameters
--- here $m_\pi$ and $p_{thr}$ --- loops no longer contribute at a single order
but at all orders higher than where they start to contribute.} calculated
in threshold kinematics --- that is neglecting the distortions from the
$NN$ final-- and initial state interaction and putting all final states at
rest.
At threshold only two amplitudes contribute, namely the one
with the nucleon pair in the final and initial state in isospin 1
(measured, e.g.,
in $pp\to pp\pi^0$) and the one where the total $NN$
isospin is changed from 1 to 0 (measured, e.g., in $pp\to d\pi^+$). 
It was found that
 the sum all loops that contain $\Delta$--excitations vanish in both channels.
This was understood, since the loops were divergent and at NLO no counter term
is allowed by chiral symmetry. On the other hand
 the nucleonic loops were individually finite. It was found that
the sum of all nucleonic loops that contribute to $pp\to pp\pi^0$ 
vanish, whereas
the  sum of those that contribute to $pp\to d\pi^+$
gave a finite answer. The resulting amplitude grows linear
with the initial momentum.
In Ref.~\cite{gep} it was pointed out that
this growth of the amplitude is
problematic: when evaluated for finite outgoing $NN$ momenta,
the transition amplitudes turned out to scale as the momentum transfer.
Especially, the amplitudes then grew linearly with the external $NN$ momenta.
As a consequence, once convoluted with the $NN$ wave functions, a large
sensitivity to those was found, in conflict with general requirements from
field theory.  The solution
to this puzzle was presented in Ref. \cite{lensky2} and will be reported 
now.

\begin{figure}[t!]
\begin{center}
\psfrag{xx1}{$(m_\pi,\vec 0)$}
\psfrag{xx2}{$(l_0,\vec l)$}
\psfrag{yy1}{$(E+l_0-m_\pi,\vec p+\vec l)$}
\psfrag{yy2}{$(E,\vec p)$}
\psfrag{VV}{\Large $V_{\pi\pi NN}=$}
\epsfig{file=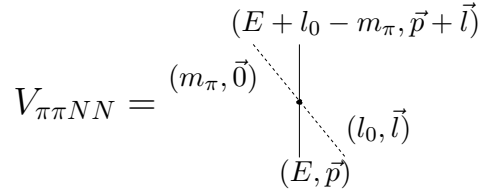, height=2.2cm}
\end{center}
\caption{The $\pi N\to \pi N$ transition
vertex: definition of kinematic variables as used in the text.}
\label{vpipinn}
\end{figure}

The observation central to the analysis is that the leading $\pi N\to
\pi N$ transition vertex, as it appears in Fig. \ref{tree}$a$, is energy
dependent. Using the notation of Fig. \ref{vpipinn} its momentum and
energy dependent part may be written as \cite{ulfs}
\begin{eqnarray}
V_{\pi\pi NN}&=&
%\frac{1}{4f_\pi^2}\epsilon^{abc}\tau^c\left(
l_0{+}m_\pi{-}\frac{\vec l\cdot(2\vec p+\vec l)}{2M_N} \nonumber \\
&=&%\frac{1}{4f_\pi^2}\epsilon^{abc}\tau^c\left(
\underbrace{{2m_\pi}}_{\mbox{on-shell}}{+}\underbrace{{\left(l_0{-}m_\pi{+}E{-}
\frac{(\vec l+\vec p)^2}{2M_N}\right)}}_{(E'-H_0)=(S')^{-1}}{-}\underbrace{{\left(E{-}\frac{\vec p\,
    ^2}{2M_N}\right)}}_{(E-H_0)=S^{-1}} \ .
\label{pipivert}
\end{eqnarray}
For simplicity we skipped the isospin part of the amplitude. The first term in
the last line denotes the transition in on--shell kinematics, the second the
inverse of the outgoing nucleon propagator and third the inverse of the
incoming nucleon propagator.  First of all we observe that for on--shell
incoming and outgoing nucleons, the the $\pi N\to \pi N$ transition vertex
takes its on--shell value $2m_\pi$ --- even if the incoming pion is
off--shell, as it is for diagram $(a)$ of Fig. \ref{tree}.  This is in
contrast to standard phenomenological treatments~\cite{kur}, where $l_0$ was
identified with $m_\pi/2$ --- the energy transfer in on--shell kinematics ---
and the recoil terms were not considered. Note, since $p_{thr}^2/M_N=m_\pi$
the recoil terms are to be kept.

\begin{figure}[t!]
\begin{center}
\epsfig{file=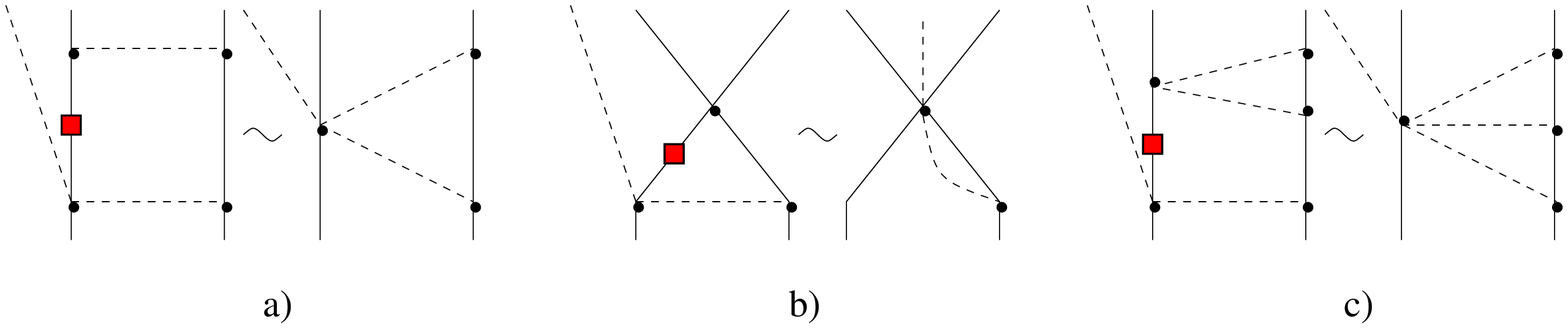, height=2.4cm}
\end{center}
\caption{Induced irreducible topologies, when the
off--shell terms of Eq. (\ref{pipivert}) hit the $NN$ potential
in the final state. The filled box on the nucleon line denotes
the propagator canceled by the off--shell part of the vertex. }
\label{irred}
\end{figure}

A second consequence of Eq. (\ref{pipivert}) is even more interesting: when
the $\pi N\to \pi N$ vertex gets convoluted with $NN$ wave functions, only the
first term leads to a reducible diagram. The second and third term, however,
lead to irreducible contributions, since one of the nucleon propagators gets
canceled.  This is illustrated in Fig. \ref{irred}, where those induced
topologies are shown that appear, when one of the nucleon propagators is
canceled (marked by the filled box) in the convolution of typical diagrams of
the $NN$ potential with the $NN\to NN\pi$ transition operator. Power counting
gives that diagram $(b)$ and $(c)$ appear only at order N$^4$LO and N$^3$LO,
respectively. However, diagram $(a)$ starts to contribute at NLO and it was
found in Ref. \cite{lensky2} that those induced irreducible contributions
cancel the finite remainder of the NLO loops in the $pp\to d\pi^+$ channel.
Thus, up to NLO only the diagrams of Fig. \ref{tree} contribute to $pp\to
d\pi^+$, with the rule that the $\pi N\to \pi N$ vertex is put on--shell.

\begin{figure}[t!]
\begin{center}
\psfig{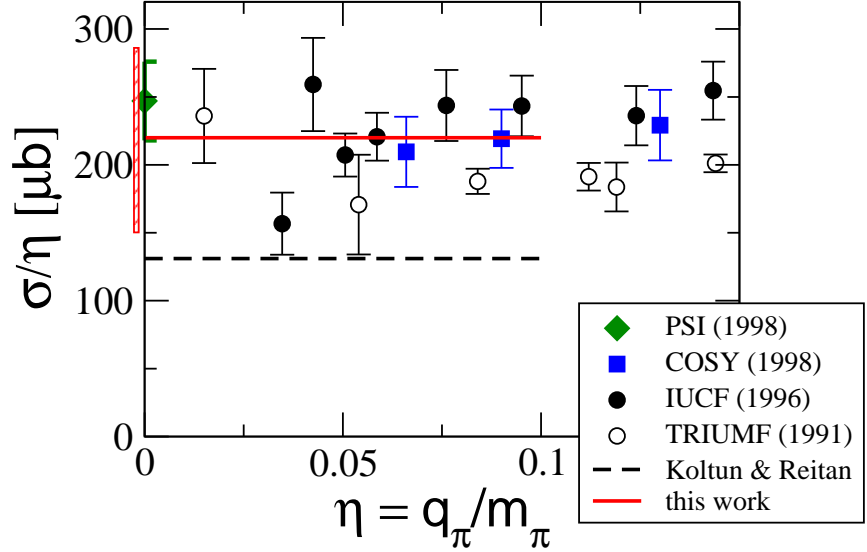}
\end{center}
\caption{Comparison of the results of Ref.~\cite{lensky2}
 to experimental data for $NN\to d\pi$. 
  The dashed line corresponds to the model of Koltun and Reitan~\cite{kur},
  whereas the solid line is the result of the ChPT calculation of
  Ref.~\cite{lensky2}.  The estimated theoretical uncertainty (see text) is
  illustrated by the narrow box.  The data is from Refs. \cite{dpidata1} (open
  circles), \cite{dpidata2} (filled circles) and \cite{dpidata3} (filled
  squares). The first data set shows twice the cross section for $pn\to
  d\pi^0$ and the other two the cross section for $pp\to d\pi^+$.}
\label{NNpi_result}
\end{figure}

The result found in Ref. \cite{lensky2} is shown in Fig. \ref{NNpi_result}
as the solid line, where
the total cross section (normalized by the energy dependence of phase space) 
is plotted against the normalized pion momentum. The dashed line
is the result of the model by Koltun and Reitan~\cite{kur}, as described
above. The data sets are from TRIUMF~\cite{dpidata1}, IUCF~\cite{dpidata2},
and COSY~\cite{dpidata3}.

\section{Comparison to phenomenological works}

All recent phenomenological calculations for $NN\to NN\pi$ add additional
diagrams to the model of Ref.~\cite{kur}. Here we will focus only on $pp\to
d\pi^+$. Phenomenological calculations for this reaction in near threshold
kinematics are given, e.g., in Ref.~\cite{jouni} and Ref.~\cite{roleofdel}.
In both works in addition to the diagrams of Ref.~\cite{kur} some
$\Delta$--loops as well as additional short range contributions are included
--- heavy meson exchanges for the former and off--shell $\pi N$
scattering\footnote{That those are also short range contributions is discussed
  in Ref.~\cite{report}.}  for the latter. Based on this the cross section for
$pp\to d\pi^+$ is overestimated near threshold.  How can we interpret this
discrepancy in light of the discussion above?

First of all, the NLO parts of the $\Delta$--loops cancel, as was
shown already in Ref.~\cite{withnorbert}. However, in both 
Refs.~\cite{jouni,roleofdel} only one of these diagrams was included
and, especially for Ref. \cite{jouni}, gave a significant contribution.
In addition, in the effective field theory short ranged operators
start to contribute only at N$^2$LO. The only diagram of
those NLO loops shown in Fig.~\ref{NLOdiags} that is effectively
included in Ref.~\cite{roleofdel} is the fourth, since the
pion loop there can be regarded as part of the $\pi N\to \pi N$
transition $T$--matrix. However, as described, the contribution
of this diagram gets canceled by the others shown in Fig.~\ref{NLOdiags}
and the induced irreducible pieces described above. Therefore, the
physics that enhances the cross section compared to the work of Ref.~\cite{kur}
in Refs.~\cite{jouni,roleofdel} is completely different to that
of Ref.~\cite{lensky2}. However, only the last one is field theoretically
consistent as explained in the previous section.

The natural question that arises is that for observable consequences. 
As explained, in the effective field theory calculation the
near threshold cross section for $pp\to d\pi^+$ is basically given
by a long--ranged pion exchange diagram, whereas the 
phenomenological calculations rely on short ranged operators with
respect to the $NN$ system. Obviously those observables are
sensitive to this difference that get prominent contributions
from higher partial waves in the final $NN$ system. We therefore
need to look at the reaction $pp\to pn\pi^+$. Unfortunately, the total
cross section for this
reaction is largely saturated by $NN$ $S$--waves in the final state (see, e.g.,
Fig.~17 in Ref.~\cite{report}). On the other hand, linear combinations of
 double polarization observables allow one to remove the prominent 
components and the sub-leading amplitudes should be visible. We
therefore expect from the above considerations that the phenomenological
calculations give good results for polarization observables for $pp\to d\pi^+$,
whereas there should be deviations for some of those for $pp\to pn\pi^+$.
Predictions for these observables were presented in Ref.~\cite{polobs} and
indeed the $\pi^+$ observables with the deuteron in the final state,
reported in Ref.~\cite{iucf2},
are described well whereas there are discrepancies for the $pn$ final state
(see Fig.~24 of Ref.~\cite{report}). For the corresponding data see
Ref.~\cite{iucf3}.

It remains to be seen how well the same data can be described in the
effective field theory framework. Up to NNLO the number of counter terms
is quite low: there are two counter terms for pion $s$--waves, that
can be arranged to contribute to $pp\to pp\pi^0$ and $pp\to d\pi^+$
individually, and then there is one counter term for pion $p$--waves,
that contributes only to a small amplitude in charged pion
production~\cite{ch3body}. On the other hand there is a huge
amount of even  double polarized data available~\cite{iucf2,iucf3,iucf1} --- and
there is more to come especially for $pn\to pp\pi^-$~\cite{ankeprop}.

\section{Dispersive corrections to $a_{\pi d}$}
\label{pid}

Let us come back to $\pi d$ scattering at threshold. The corresponding
scattering length was presented in Eq. (\ref{exp}). In the introductory
sections we exclusively focused on the real part. However, now
we are in the position to also discuss the imaginary part,
which is closely linked to the reaction $NN\to d\pi$
through unitarity and detailed balance. One may write
\begin{equation}
4\pi \mbox{Im}(a_{\pi d})=\lim_{q\to 0}q\left\{
\sigma(\pi d\to NN)+\sigma(\pi d\to \gamma NN)\right\} \ ,
\label{opttheo}
\end{equation}
where $q$ denotes the relative momentum of the initial $\pi d$ pair.
The ratio $R=\lim_{q\to 0}\left(\sigma(\pi d\to NN)/\sigma(\pi d\to
\gamma NN)\right)$ was measured to be $2.83\pm 0.04$
\cite{highland}. At low energies diagrams that lead to a sizable
imaginary part of some amplitude are expected to also contribute
significantly to its real part.
% due to
%the so--called principal value part of the corresponding
%integral.
 Those contributions are called dispersive corrections.  As a
first estimate Br\"uckner speculated that the real and imaginary part
of these contributions should be of the same order of magnitude
\cite{brueck}. This expectation was confirmed within Faddeev
calculations in Refs. \cite{at}.  Given the high accuracy of the
measurement and the size of the imaginary part of the scattering
length, another critical look at this result is called for as already
stressed in Refs. \cite{tle,bk}.  A consistent calculation is only
possible within a well defined effective field theory --- the first
calculation of this kind was presented in 
Ref.~\cite{apiddisp} and is briefly sketched here.

\begin{figure}[t!]
\begin{center}
\psfig{file=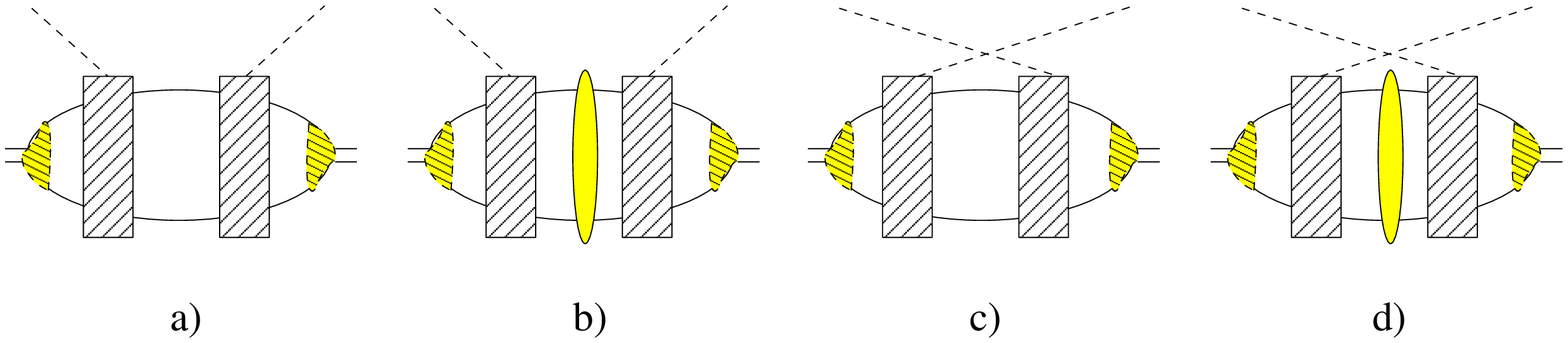,width=4.in}
\end{center}
\caption{Dispersive corrections to the $\pi d$ scattering length.}
\label{disp}
\end{figure}

To identify the diagrams that are to contribute we first need to specify
what we mean by a dispersive correction.
We define dispersive corrections as contributions from diagrams with
an intermediate state that contains only nucleons, photons and at most
real pions. Therefore, all the diagrams shown in Fig.~\ref{disp}
are included in our work. On the other hand, all diagrams
that, e.g., have Delta excitations in the intermediate state
do not qualify as dispersive corrections, although they
might give significant contributions~\cite{doeringoset}.

The hatched blocks in the diagrams of Fig.~\ref{disp} refer
to the relevant transition operators for the reaction $NN\to NN\pi$
depicted in Fig.~\ref{tree}. Also in the kinematics of relevance
here the $\pi N\to \pi N$ transitions are to be taken with
their on--shell value $2m_\pi$. Using the CD--Bonn potential~\cite{cdbonn}
for the $NN$ distortions we found for the dispersive
correction from the purely hadronic transition
\begin{equation}
a_{\pi d}^{disp}=(-6.3+2+3.1-0.4)\times 10^{-3} \, m_\pi^{-1}
=-1.6\times 10^{-3} \, m_\pi^{-1} \ ,
\end{equation}
where the numbers in the first bracket are the individual results for the
diagrams shown in Fig.~\ref{disp}, in order.  Note that the diagrams with
intermediate $NN$ interactions and the crossed ones (diagram $(c)$ and $(d)$),
neither of them included in most of the previous calculations, give
significant contributions.  The latter finding might come as a surprise on the
first glance, however, please recall that in the chiral limit all four
diagrams of Fig.~\ref{disp} are kinematically identical and chiral
perturbation theory is a systematic expansion around exactly this point. Thus,
as a result we find that the dispersive corrections to the $\pi d$ scattering
length are of the order of 6 \% of the real part of the scattering length.
This number is fully in line with the expectations from power counting, which
predicted a relative suppression of the dispersive corrections compared to the
leading double scattering term --- diagram $(b)$ of Fig. \ref{leadingpid} ---
of the order of $(m_\pi/M_N)^{3/2}\sim 5$ \%. Note that the same calculation
gave very nice agreement for the corresponding imaginary part~\cite{apiddisp}.

In Ref.~\cite{apiddisp} also the electro--magnetic contribution to the
dispersive correction was calculated. It turned out that the contribution to
the real part was tiny --- $-0.1\times 10^{-3} \, m_\pi^{-1}$ --- while the
sizable experimental value for the imaginary part (c.f.  Eqs. (\ref{exp}) and
\ref{opttheo}) was described well.

To get a reliable estimate of the uncertainty of the calculation just
presented a NNLO calculation is necessary. At that order a counter
term appears for pions at rest that can be fixed from $NN\to NN\pi$,
as indicated above. For now we can only present a conservative estimate for
the uncertainty by using the uncertainty of order $2\, m_\pi/M_N$ one
has for, e.g., the sum of all direct diagrams to derive a $\Delta
a_{\pi d}^{disp}$ of around $1.4\times 10^{-3} \, m_\pi^{-1}$, which
corresponds to about 6\% of $\mbox{Re}\left(a_{\pi
d}^{\mbox{exp}}\right)$. However, given that the operators that
contribute to both direct and crossed diagrams are almost the same
 and that part of the mentioned cancellations is a
direct consequence of kinematics, this number for $\Delta a_{\pi
d}^{disp}$ is probably too large.

In Ref.~\cite{apiddisp} a detailed comparison to previous
works is given. Differences in the values found for the
dispersive corrections were traced to the incomplete sets
of diagrams included in those phenomenological studies.

\section{Summary and Outlook for pion reactions}

In the lectures recent progress in 
our understanding of elastic and inelastic
pion reactions on the two--nucleon system was presented.
The central reaction discussed was $\pi d$ scattering
at threshold. Arguments were given that in the years
to come one should be able to calculate the $\pi d$ scattering
length with sufficient accuracy to use the reaction
as one of the prime sources for the isoscalar scattering
length $a_+$. To reach this not only significant progress
was necessary for the coherent $\pi d$ scattering but
also the reactions $NN\to NN\pi$ need to understood.
In the future also the role of isospin violation on 
$\pi d$ scattering needs to be investigated further as stressed in
Ref.~\cite{MRR}.

The process $NN\to NN\pi$ is a puzzle already since more
than a decade. Given the progress presented above we have
now reason to believe that this puzzle will be solved soon.
This mentioned results could only be found, because a consistent
effective field theory was used. For example, the potential problem
with the transition operators of Ref.~\cite{withnorbert},
pointed at in Ref.~\cite{gep}, would
always be hidden in phenomenological calculations, since the
form factors routinely used there always lead to finite, well behaved
 amplitudes.
The very large number of observables available for
the reactions $NN\to NN\pi$ will provide a non--trivial test
to the approach described.

Once the scheme is established, the same field theory can
be used to analyze the isospin violating observables measured
in $pn\to d\pi^0$~\cite{opper} and $dd\to \alpha\pi^0$~\cite{ed}.
First steps in this direction were already done in Ref.~\cite{knm}
for the former and in Refs.~\cite{csb1,csb2} for the latter.

In the lectures also the reaction $\gamma d\to\pi^+ nn$ was discussed.
Not only gave those studies a further confirmation that we understand
the few body dynamics well within ChPT, but it also promises to become
an ideal reaction for the extraction of the $nn$ scattering length with
high accuracy. The corresponding measurements could be performed at
HIGS~\cite{aaronpriv}.

\section{Some remarks on strangeness production}
\label{YN}

\begin{figure}[t!]
\begin{center}
  \psfig{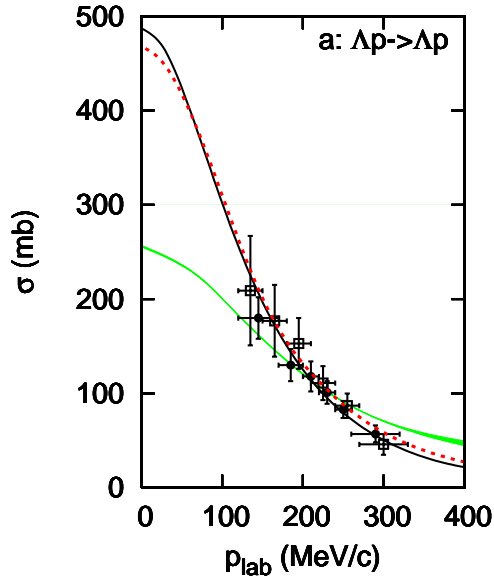}
\end{center}
\caption{Comparison of different variants
for the $\Lambda N$ interaction to the available 
data at low energies. The solid line, dashed line, and the
light area are the results of Refs.~\cite{rijken,Hai05,henk}, in order.
The data are from Refs.~\cite{data1,data1b}.}
\label{lnscl}
\end{figure}

As described in the previous sections a lot is already known
about the properties and dynamics of systems composed of 
light quarks only. However, much less is known about
the scattering of
systems with strangeness, especially for low energies.
The reason is of experimental nature: the lifetime of
particles with strangeness is typically too short to
allow for secondary scatterings at low energies, necessary
to get information on low energy scattering.

Here we will focus on the hyperon--nucleon ($YN$) system. 
The poor status of our information on the $YN$ interaction is
most obviously reflected in the present knowledge of the $\Lambda N$ 
scattering lengths. Attempts in the 1960's to pin down the low
energy parameters for the $S$-waves led to results that were afflicted by rather 
large uncertainties \cite{data1,data1b}. In Ref. \cite{data1b}
the following values are given for the singlet scattering length $a_s$ and the
triplet scattering length $a_t$
\begin{equation}
a_s=-1.8\left\{{+2.3 \atop -4.2}\right. \ \mbox{fm and } 
a_t=-1.6\left\{{+1.1 \atop -0.8}\right. \ \mbox{fm} ,
\end{equation}
where the errors are strongly correlated. The situation of the corresponding
effective ranges is even worse: for both spin states values between 0 and 16 fm
are allowed by the data.
Later, the application of microscopic models 
for the extrapolation of the data to the threshold, was hardly more successful
to pin down the low energy parameters. 
For example, in 
Ref. \cite{rijken} one can find six different models that equally well
describe the available data but whose ($S$-wave) scattering lengths 
range from -0.7 to -2.6 fm in the singlet channel and from -1.7 to -2.15 fm 
in the triplet channel. To illustrate this point in Fig. \ref{lnscl} we show
a comparison of model f of Ref.~\cite{rijken} (dark solid line), the J\"ulich '04
model~\cite{Hai05} (dashed curve), and the result from the recent effective
field
theory approach of Ref.~\cite{henk} to the world data. 

The natural alternative to scattering experiments are production reactions.
However, the central insights of the previous sections were that only
within a consistent field theory reliable calculations can be
performed for the reactions under considerations. On the other
hand, as stressed in the introduction, any strangeness production
reaction involves momenta that do not allow for an expansion
along the lines just discussed, since the corresponding expansion parameter
in this case would be larger than $1/2$. Does this mean that one can
learn nothing from a study of strangeness production off two nucleon
systems?

Not at all. For one thing, an investigation of baryon and meson
resonances does not need any detailed knowledge on the production
mechanism --- most of the relevant information is contained in
the Dalitz plots --- and the comment of the previous paragraph
applies only to this part. But one can learn even more
from the production reactions by using that the momentum transfer
in those reactions are large. Since this leads to an effectively
point--like production operator, one may employ dispersion theory
to relate invariant mass spectra of production reactions to
elastic scattering data. Obviously, for this no knowledge on the
production operator is necessary whatsoever.
Then, in contrast to above, one works with
the typical outgoing relative momentum in units of the momentum transfer
as
 expansion parameter.

 In the remainder of this text I will only focus on how to extract scattering
 parameters from production reactions. One aspect of spectroscopy, also
 discussed in the lectures, namely that of scalar mesons, was described in the
 recent conference proceeding~\cite{qnp} in very much detail and it will not
 be repeated here.

The use of dispersion theory was very common in the 50s and the basis for the
study to be described now was worked out already then~\cite{disptheo}.  A
controlled method of extraction of scattering lengths from production
reactions opens up the opportunity to measure scattering parameters also for
unstable states. As an example we will discuss the option to measure the
$\Lambda N$ scattering lengths from $NN\to K\Lambda N$ and $\gamma d\to
K\Lambda N$. In Refs.~\cite{mitachot1,mitachot2,mitachot3} it was shown, how
to derive an integral representation of the scattering length of a pair of
outgoing particles (here we show the formula relevant for neutral final
states, as is relevant for $\Lambda N$; in the presence of Coulomb
interactions the equation has to be modified --- see Ref.~\cite{mitachot2}):
\begin{eqnarray}
\nonumber
a_S&=&\lim_{{m}^2\to m_0^2}\frac1{2\pi}\left(\frac{m_\Lambda+m_N}
{\sqrt{m_\Lambda m_N}}\right){\bf P}
\int_{m_0^2}^{m_{max}^2}dm' \, ^2
\sqrt{\frac{m_{max}^2-{m}^2}{m_{max}^2-m' \, ^2}}\\
& & \qquad \qquad \times \ 
\frac1{\sqrt{m' \, ^2-m_0^2} \ (m' \, ^2-{m}^2)}
\log{\left\{\frac{1}{p'}\left(\frac{d^2\sigma_S}{dm' \, ^2dt}\right)\right\}}
\ ,
\label{final}
\end{eqnarray}
where $\sigma_S$ denotes the spin cross section for the production of a
$\Lambda$--nucleon pair with invariant mass $m' \, ^2$---corresponding to a
relative momentum $p'$---and total spin $S$. In addition $t=(p_1-p_{K^+})^2$, 
with $p_1$
being the beam momentum, $m_0^2=(m_\Lambda+m_N)^2$, where $m_\Lambda$ ($m_N$)
denotes the mass of the Lambda hyperon (nucleon), and $m_{max}$ is some
suitably chosen cutoff in the mass integration.  
In Ref.~\cite{mitachot1} it was shown that  it
is sufficient to include relative energies of the final $\Lambda N$ system of
at most 40 MeV in the range of integration to get accurate results.  $\bf P$
denotes that the principal value of the integral is to be used and the limit
has to be taken from above.

\begin{figure}[t!]
\begin{center}
  \psfig{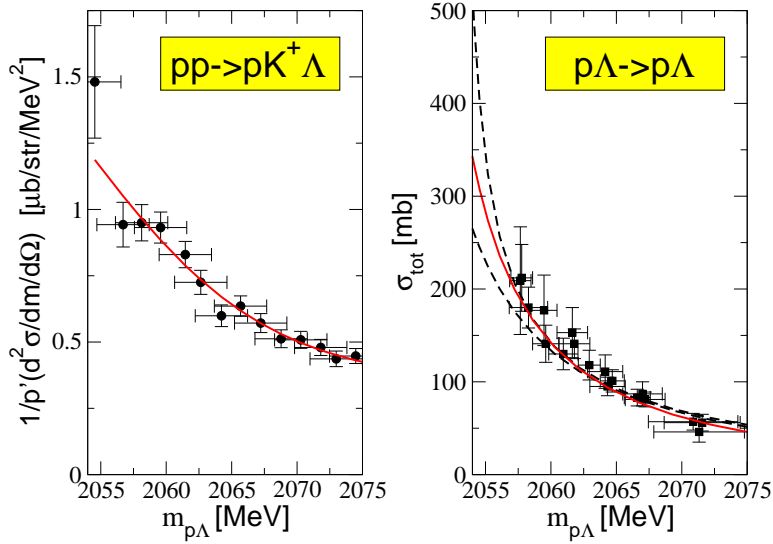}
\end{center}
\caption{Illustration of the uncertainties of extraction
of scattering parameters from scattering data (left panel), where
an extrapolation is necessary, and from production data (right panel),
where the data needs to be interpolated.
The data was taken from Refs.~\cite{data1b,sechi} for the right panel
and from Ref.~\cite{saclay} for the left one.}
\label{prodvsscat}
\end{figure}

The formula as given is applicable if there are no significant
effects from crossed channels --- this can be monitored
by a Dalitz plot analysis --- and only a single partial wave
contributes. With respect to the angular momentum this can
be achieved by a proper choice of $m_{max}$. In order to
select a single spin state for the outgoing two--particle
system polarization observables are necessary. In
Refs.~\cite{mitachot1,mitachot3}
the relevant observables are identified for the reactions 
$NN\to NKY$ and $\gamma d\to K\Lambda N$, respectively.
The former class of reactions can be measured 
at COSY~\cite{ankeprop,tofprop} and the latter
either at J-Lab~\cite{jlab}, MAMI~\cite{mami}, or ELSA~\cite{elsa}.

The essential advantages of the use of Eq.~(\ref{final})
to extract the scattering lengths of unstable particles
compared to a determination  from scattering experiments 
are, e.g., for $\Lambda N$ scattering:
\begin{itemize}
\item Instead of an extrapolation of data,
the scattering length is found from an interpolation
of an invariant mass spectrum, which is theoretically
much better controlled. This is illustrated in Fig.~\ref{prodvsscat}.
\item The integral representation gives a result for 
the scattering length without any assumption on the
energy dependence of the hyperon--nucleon interaction.
This opens the possibility to fix the scattering lengths
from production reactions and then use scattering data
to fix, e.g., the effective range.
\item Since Eq.~(\ref{final}) is derived from a 
dispersion integral, a controlled error estimate
is possible. A systematic study revealed that
for the kinematics relevant for the production
of the $\Lambda N$ system at low energies, the
uncertainty of  Eq.~(\ref{final}) was found to
be $0.5$ fm~\cite{mitachot1,mitachot3}. Sources
of this uncertainty are the possible energy dependence
of the production operator, the influence of the 
upper limit of integration, and the possible influence
of crossed channel effects.
\end{itemize}

\section{Summary}

In these proceedings various reactions on few nucleon
systems were discussed. It was demonstrated that due
to significant advances in the technologies of effective
field theories, high precision calculations became 
possible for hadronic reactions even on few--nucleon systems.
At the same time a clear connection to QCD is provided.

For the production of more heavy systems, like those that contain
the strangeness degree of freedom, off two nucleons no effective field theory
is developed yet and therefore many reactions are still being
analyzed using models. However, also for this class of reactions
some aspects can be analyzed in a model independent way. The
$\Lambda N$ scattering lengths were discussed as an example.

We are now in a phase, were both theory and experiment are advanced
sufficiently that we should understand much better how QCD influences
low energy nuclear dynamics in the upcoming years. Especially the
symmetry breaking sector --- violation of isospin as well as
violation of the flavor $SU(3)$ --- promises deep insights
into the mechanisms of strong interactions.

\section*{Acknowledgments}
I thank the organizers for a superb job and V. Lensky, V. Baru,
A. Gasparyan,
J.~Haidenbauer, A. Kudryavtsev, and U.-G. Mei\ss ner for a very
fruitful collaboration that lead to the results presented.  
%This
%research is part of the EU Integrated Infrastructure Initiative Hadron
%Physics Project under contract number RII3-CT-2004-506078, and was
%supported also by the DFG-RFBR grant no. 05-02-04012 (436 RUS
%113/820/0-1(R)) and the DFG SFB/TR 16 "Subnuclear Structure of
%Matter".  


\begin{thebibliography}{99}

\frenchspacing

\bibitem{gilberto}
  G.~Colangelo, J.~Gasser and H.~Leutwyler,
  %``pi pi scattering,''
  Nucl.\ Phys.\  B {\bf 603}, 125 (2001)
  [arXiv:hep-ph/0103088].


\bibitem{ulfs}
V.~Bernard, N.~Kaiser and U.-G.~Mei\ss ner,
%``Chiral dynamics in nucleons and nuclei,''
Int. \ J. \ Mod. \ Phys. \ E {\bf 4} (1995) 193. 

\bibitem{nadiaq3}
  N.~Fettes and U.-G.~Mei\ss ner,
  %``Complete analysis of pion nucleon scattering in chiral perturbation  theory
  %to third order,''
  Nucl.\ Phys.\  A {\bf 693}, 693 (2001)
  [arXiv:hep-ph/0101030].


\bibitem{bastian}
B. Kubis, these proceedings.

\bibitem{birapaulo}
 P.~F.~Bedaque and U.~van Kolck,
  %``Effective field theory for few-nucleon systems,''
  Ann.\ Rev.\ Nucl.\ Part.\ Sci.\  {\bf 52} (2002) 339
  [arXiv:nucl-th/0203055].


\bibitem{evgenirev}
  E.~Epelbaum,
  %``Few-nucleon forces and systems in chiral effective field theory,''
  Prog.\ Part.\ Nucl.\ Phys.\  {\bf 57} (2006) 654
  [arXiv:nucl-th/0509032].

\bibitem{machleidt}
R.~Machleidt, 
these proceedings.

\bibitem{wein}
S.~Weinberg,
%``Three Body Interactions Among Nucleons And Pions,''
Phys.\ Lett.\ B {\bf 295} (1992) 114.

\bibitem{beane}
S.~R.~Beane, V.~Bernard, E.~Epelbaum, U.-G.~Mei{\ss}ner and D.~R.~Phillips,
 %``The S-wave pion nucleon scattering lengths from pionic atoms using
%effective field theory,''
Nucl.\ Phys.\ A {\bf 720} (2003) 399.
[arXiv:hep-ph/0206219].

\bibitem{kbl}S.~R.~Beane, V.~Bernard, T.~S.~H.~Lee, U.-G.~Mei{\ss}ner and U.~van Kolck,
  %``Neutral pion photoproduction on deuterium in baryon chiral perturbation
  %theory to order q**4,''
  Nucl.\ Phys.\ A {\bf 618}, 381 (1997)
  [arXiv:hep-ph/9702226].

\bibitem{krebs} H.~Krebs, V.~Bernard and U.-G.~Mei\ss ner, 
%``Improved analysis  of neutral pion electroproduction off deuterium in 
%chiral perturbation theory,'' 
Eur.\ Phys.\ J.\ A {\bf 22} (2004) 503 [arXiv:nucl-th/0405006].

\bibitem{baru}
  V.~Baru, J.~Haidenbauer, C.~Hanhart and J.~A.~Niskanen,
  %``New parameterization of the trinucleon wavefunction and its application  to
  %the pi He-3 scattering length,''
  Eur.\ Phys.\ J.\ A {\bf 16}, 437 (2003)
  [arXiv:nucl-th/0207040].

\bibitem{garde}
  A.~Gardestig and D.~R.~Phillips,
  % ``Using chiral perturbation theory to extract the neutron neutron  scattering
  %length from pi- d --> n n gamma,''
  Phys.\ Rev.\ C {\bf 73} (2006) 014002
  [arXiv:nucl-th/0501049];  A.~Gardestig and D.~R.~Phillips,
  %``How low-energy weak reactions can constrain three-nucleon forces and  the
  %neutron neutron scattering length,''
  arXiv:nucl-th/0603045.


\bibitem{lensky}
  V.~Lensky, V.~Baru, J.~Haidenbauer, C.~Hanhart, A.~E.~Kudryavtsev and U.-G.~Mei\ss ner,
  %``Precision calculation of gamma d --> pi+ n n within chiral perturbation
  %theory,''
  Eur.\ Phys.\ J.\ A {\bf 26}, 107 (2005)
  [arXiv:nucl-th/0505039].




\bibitem{recoils}
  V.~Baru, C.~Hanhart, A.~E.~Kudryavtsev and U.-G.~Mei\ss ner,
  %``The role of the nucleon recoil in low-energy meson nucleus reactions,''
  Phys.\ Lett.\ B {\bf 589}, 118 (2004)
  [arXiv:nucl-th/0402027].
 
\bibitem{park}
B.Y. Park et al., Phys. Rev. C {\bf 53} (1996) 1519
[arXiv:nucl-th/9512023].

\bibitem{unserd}
  C.~Hanhart, J.~Haidenbauer, M.~Hoffmann, U.-G.~Mei{\ss}ner and J.~Speth,
  %``The reactions p p $\to$ p p pi0 and p p $\to$ d pi+ at threshold: The role
  %of the isoscalar pi N scattering amplitude,''
  Phys.\ Lett.\ B {\bf 424} (1998) 8
  [arXiv:nucl-th/9707029].

\bibitem{haidenbauermachner} 
  H.~Machner and J.~Haidenbauer,
  %``Meson production close to threshold,''
  J.\ Phys.\ G {\bf 25}, R231 (1999).
 

\bibitem{dmit} V. Dmitra\v sinovi\' c, K. Kubodera, F. Myhrer and  T. Sato,
{Phys. Lett.} B {\bf  465} (1999) 43   [arXiv:nucl-th/9902048].


\bibitem{ando}
  S.~I.~Ando, T.~S.~Park and D.~P.~Min,
  %``Threshold p p $\to$ p p pi0 up to one-loop accuracy,''
  Phys.\ Lett.\ B {\bf 509} (2001) 253
  [arXiv:nucl-th/0003004]. 


\bibitem{bira1} T.D. Cohen, J.L. Friar, G.A. Miller and U. van Kolck, 
{Phys. Rev.} C {\bf 53} (1996) 2661 [arXiv:nucl-th/9512036].

\bibitem{rocha}
  C.~da Rocha, G.~Miller and U.~van Kolck,
  %``The N N $\to$ N N pi+ reaction near threshold in a chiral power counting
  %approach,''
  Phys.\ Rev.\ C {\bf 61} (2000) 034613
  [arXiv:nucl-th/9904031].


\bibitem{ch3body}
 C. Hanhart, U. van Kolck, and
  G.A. Miller, Phys. Rev. Lett. {\bf 85} (2000) 2905 [arXiv:nucl-th/0004033].

\bibitem{report}
C. Hanhart, Phys. Rep. \textbf{397} (2004) 155 [arXiv:hep-ph/0311341].


\bibitem{withnorbert}
C.~Hanhart and N.~Kaiser,
%``Complete next-to-leading order calculation for pion production in nucleon
%nucleon collisions at threshold,''
Phys.\ Rev.\ C {\bf 66} (2002) 054005 [arXiv:nucl-th/0208050].


\bibitem{wallacephillips}
  D.~R.~Phillips, S.~J.~Wallace and N.~K.~Devine,
  % ``Electron deuteron scattering in the equal-time formalism: Beyond the
  %impulse approximation,''
  Phys.\ Rev.\ C {\bf 72} (2005) 0140061.
  [arXiv:nucl-th/0411092].

\bibitem{PSI1}
 P.~Hauser {\it et al.},
  %``New precision measurement of the pionic deuterium s-wave strong
  %interaction parameters,''
  Phys.\ Rev.\ C {\bf 58} (1998) 1869;
  D.~Chatellard {\it et al.},
  %``X-ray spectroscopy of the pionic deuterium atom,''
  Nucl.\ Phys.\ A {\bf 625} (1997) 855.

\bibitem{detlev}
D. Gotta et al., PSI experiment R-06.03; D. Gotta, private communication.

\bibitem{BBEMP}
S.~R.~Beane, V.~Bernard, E.~Epelbaum, U.-G.~Mei{\ss}ner and D.~R.~Phillips,
 %``The S-wave pion nucleon scattering lengths from pionic atoms using
%effective field theory,''
Nucl.\ Phys.\ A {\bf 720} (2003) 399
[arXiv:hep-ph/0206219].
\bibitem{rus}
  U.-G.~Mei{\ss}ner, U.~Raha and A.~Rusetsky,
  %``The pion nucleon scattering lengths from pionic deuterium,''
  Eur.\ Phys.\ J.\ C {\bf 41} (2005) 213
  [arXiv:nucl-th/0501073].
\bibitem{doeringoset}
  M.~D\"oring, E.~Oset and M.~J.~Vicente Vacas,
  %``S-wave pion nucleon scattering length from pi N, pionic hydrogen and
  %deuteron data,''
  Phys.\ Rev.\ C {\bf 70} (2004) 045203
  [arXiv:nucl-th/0402086].
\bibitem{arriola}
 M.~Pavon Valderrama and E.~Ruiz Arriola,
  %``Renormalization of the deuteron with one pion exchange,''
  Phys.\ Rev.\ C {\bf 72} (2005) 054002
  [arXiv:nucl-th/0504067];
  M.~P.~Valderrama and E.~R.~Arriola,
  %``Pion Deuteron Scattering and Chiral Expansions,''
  arXiv:nucl-th/0605078.
\bibitem{mitandreas}
  A.~Nogga and C.~Hanhart,
  %``Can one extract the pi neutron scattering length from pi deuteron
  %scattering?,''
  Phys.\ Lett.\ B {\bf 634} (2006) 210
  [arXiv:nucl-th/0511011].
\bibitem{danielneu}
  L.~Platter and D.~R.~Phillips,
  %``Deuteron matrix elements in chiral effective theory at leading order,''
  arXiv:nucl-th/0605024.
\bibitem{MRR}
U.-G.~Mei{\ss}ner, U.~Raha and A.~Rusetsky, 
%``Isospin-breaking corrections in the pion deuteron scattering length,''
Phys. Lett.B  {\bf 639} (2006) 478
  [arXiv:nucl-th/0512035].
  %%CITATION = NUCL-TH 0512035;%%
\bibitem{al}
  A.~W.~Thomas and R.~H.~Landau,
  %``Pion - Deuteron And Pion - Nucleus Scattering: A Review,''
  Phys.\ Rept.\  {\bf 58} (1980) 121.

\bibitem{gries}  B.~Borasoy and H.~W.~Grie{\ss}hammer,
  %``The S wave pion deuteron scattering length in effective field theory,''
  Int. J. of Mod. Phys. {\bf E} 12, 65 (2003) [arXiv:nucl-th/0105048].
\bibitem{silasmartin}
  S.~R.~Beane and M.~J.~Savage,
  %``Pions in the pionless effective field theory,''
  Nucl.\ Phys.\ A {\bf 717} (2003) 104
  [arXiv:nucl-th/0204046].


\bibitem{robilottawilkin}
  M.~R.~Robilotta and C.~Wilkin,
  %``On The Pion - Deuteron Scattering Length,''
  J.\ Phys.\ G {\bf 4}, L115 (1978).


\bibitem{KK}
V.~M.~Kolybasov and A.~E.~Kudryavtsev,
%``Pion-Deuteron Scattering Length,''
Nucl.\ Phys.\ B {\bf 41} (1972) 510.

\bibitem{recoils2}
  V.~Lensky, V.~Baru, J.~Haidenbauer, C.~Hanhart, A.~E.~Kudryavtsev and U.-G.~Mei\ss ner,
  %``Precision calculation of gamma d --> pi+ n n within chiral perturbation
  %theory,''
  Eur.\ Phys.\ J.\  A {\bf 26}, 107 (2005)
  [arXiv:nucl-th/0505039].

\bibitem{faeldt}
G.~F\"aldt, 
%``Binding corrections and the pion-deuteron scatt. length'',
Phys. \ Scripta {\bf 16}, 81 (1977); A.~Rusetsky, private
communication.


\bibitem{laget} J. M. Laget, Phys. Rep. {\bf 69} (1981) 1. 


\bibitem{threspiprod}
V.~Bernard, N.~Kaiser and U.-G.~Mei\ss ner,
  %``Threshold pion photoproduction in chiral perturbation theory,''
  Nucl.\ Phys.\ B {\bf 383} (1992) 442.


\bibitem{heavyChPT}H. W. Fearing, T. R. Hemmert, R. Lewis, and C.~Unkmeir,
Phys.  Rev. {\bf C 62} (2000) 054006 [arXiv:hep-ph/0005213]. 


\bibitem{MIT}E. C. Booth et al., Phys. Rev.  {\bf C20} (1979) 1217. 



\bibitem{nnfrompd}
D.E. Gonzalez--Trotter et al.,  Phys. Rev. Lett. {\bf 83} (1999) 3788;
V. Huhn et al.,  Phys. Rev. Lett. {\bf 85} (2000) 1190.



\bibitem{mitandreasw}
  C.~Hanhart and A.~Wirzba,
  %``Remarks on N N -> N N pi beyond leading order,''
  arXiv:nucl-th/0703012.


\bibitem{gep}  A.~G{\aa}rdestig, talk presented at ECT* workshop 
 'Charge Symmetry
Breaking and Other Isospin Violations', Trento, June 2005;
  A.~G{\aa}rdestig, D.~R.~Phillips and C.~Elster,
  %``The near threshold N N $\to$ d pi reaction in chiral perturbation theory,''
  arXiv:nucl-th/0511042.

\bibitem{lensky2}
  V.~Lensky, V.~Baru, J.~Haidenbauer, C.~Hanhart, A.~E.~Kudryavtsev and U.-G.~Mei\ss ner,
  %``Towards a field theoretic understanding of N N --> N N pi,''
  Eur.\ Phys.\ J.\ A {\bf 27}, 37 (2006)
  [arXiv:nucl-th/0511054].


\bibitem{kur} D. Koltun and A. Reitan, 
Phys. Rev. {\bf 141} (1966) 1413.


\bibitem{dpidata1}
  D.~A.~Hutcheon {\it et al.},
  %``Measurements of N N $\to$ d pi very near threshold. 1. The n p $\to$ d pi0
  %cross-section,''
  Nucl.\ Phys.\ A {\bf 535} (1991) 618. 

\bibitem{dpidata2}
P. Heimberg  {\it et al.}, Phys. Rev. Lett. {\bf 77} (1996) 1012. 

\bibitem{dpidata3}
 M. Drochner {\it et al.},
  Nucl.\ Phys.\ A {\bf 643} (1998) 55. 

\bibitem{jouni}
  J.~A.~Niskanen,
%   ``Comment on 'Role of heavy meson exchange in near threshold N N $\to$ d
  %pi',''
  Phys.\ Rev.\ C {\bf 53}, 526 (1996)
  [arXiv:nucl-th/9502015].


\bibitem{roleofdel}
  C.~Hanhart, J.~Haidenbauer, O.~Krehl and J.~Speth,
  %``Role of the Delta isobar in the reaction N N --> N N pi near  threshold,''
  Phys.\ Lett.\ B {\bf 444} (1998) 25
  [arXiv:nucl-th/9808020].

\bibitem{polobs}
  C.~Hanhart, J.~Haidenbauer, O.~Krehl and J.~Speth,
  %``Polarization phenomena in the reaction N N --> N N pi near threshold,''
  Phys.\ Rev.\ C {\bf 61} (2000) 064008
  [arXiv:nucl-th/0002025].


\bibitem{iucf2}
B.~von Przewoski et~al.,
 Phys. \ Rev. \ C {\bf 61} (2000) 064604.

\bibitem{iucf3}
W.~W. Daehnick et~al., Phys. \ Rev. \ C {\bf 65} (2002) 024003.


\bibitem{iucf1}
H.~O. Meyer et~al., Phys. \ Rev. \ C {\bf 63} (2001) 064002.



\bibitem{ankeprop}
  A.~Kacharava {\it et al.},
  %``Spin physics from COSY to FAIR,''
  arXiv:nucl-ex/0511028.


\bibitem{highland}
V.~C.~Highland et al., Nucl.\ Phys.\ A {\bf 365} (1981) 333. 

\bibitem{brueck}
K.~Br\"uckner, Phys. Rev. {\bf 98} (1955) 769.

\bibitem{at}
I.R.~Afnan and A.W.~Thomas, Phys. Rev. C{\bf 10} (1974) 109;
 D.S.~Koltun and T.~Mizutani, Ann. Phys. (N.Y.) {\bf 109} (1978) 1.

\bibitem{tle}
T.~ E.~ O.~ Ericson, B.~ Loiseau, A.~ W.~ Thomas, Phys. Rev. C {\bf 66} (2002) 014005 [arXiv:hep-ph/0009312].
 %%Nucl.\ Phys.\ A {\bf 684} (2001) 380.

\bibitem{bk} V.Baru, A. Kudryavtsev, Phys. Atom. Nucl., {\bf 60} (1997) 1476.

\bibitem{apiddisp} V.~Lensky, V.~Baru, J.~Haidenbauer, C.~Hanhart,
   A.~E.~Kudryavtsev and U.-G.~Mei\ss ner, 
   arXiv:nucl-th/0608042; Phys. Lett. B, in print.  


\bibitem{cdbonn}
  R.~Machleidt, 
  %``The high-precision, charge-dependent Bonn nucleon-nucleon potential
  %(CD-Bonn),''
  Phys.\ Rev.\ C {\bf 63} (2001) 024001
  [arXiv:nucl-th/0006014].

\bibitem{opper}
A.~K.~Opper {\it et al.},
Phys.\ Rev.\ Lett.\  {\bf 91} (2003) 212302  [arXiv:nucl-ex/0306027]. 

\bibitem{ed}
E.~J.~Stephenson {\it et al.},
Phys.\ Rev.\ Lett.\  {\bf 91} (2003) 142302  [arXiv:nucl-ex/0305032]. 

\bibitem{knm} 
U.~van~Kolck, J.~A.~Niskanen, and G.~A.~Miller, 
Phys.\ Lett.\ B {\bf 493}  (2000) 65 [arXiv:nucl-th/0006042]. 

\bibitem{csb1}
A.~G{\aa}rdestig {\it et al.},
Phys.\ Rev.\ C {\bf 69} (2004) 044606[arXiv:nucl-th/0402021]. 

\bibitem{csb2}
  A.~Nogga {\it et al.},
  %``Realistic few-body physics in the d d --> alpha pi0 reaction,''
  Phys.\ Lett.\  B {\bf 639}, 465 (2006)
  [arXiv:nucl-th/0602003].

\bibitem{aaronpriv}
A. Bernstein, private communication.

\bibitem{data1} 
G. Alexander et al., Phys. Lett. {\bf 19}, 715 (1966); 
B. Sechi-Zorn et al., Phys. Rev. {\bf 175}, 1735 (1968). 

%\bibitem{data2} 
%H.G. Dosch et al., Phys. Lett. {\bf 21}, 236 (1966);
%R. Engelmann et al., Phys. Lett. {\bf 21}, 587 (1966);
%F. Eisele et al., Phys. Lett. {\bf 37B}, 204 (1971).
%
%\bibitem{data3} 
%J.K. Ahn et al., Nucl. Phys. {\bf A648}, 263 (1999);
%Y. Kondo et al., Nucl. Phys. {\bf A676}, 371 (2000);
%T. Kadowaki et al., Eur. Phys. J. A {\bf 15}, 295 (2002). 

\bibitem{data1b} 
G. Alexander et al., Phys. Rev. {\bf 173}, 1452 (1968); 

\bibitem{rijken}
T.~A. Rijken, V.~G.~J. Stoks, Y.~Yamamoto, Phys. Rev. C 59 (1999) 21.


\bibitem{Hai05}
J.~Haidenbauer, U.-G. Mei{\ss}ner, Phys. Rev. C 72 (2005) 044005.


\bibitem{henk}
  H.~Polinder, J.~Haidenbauer and U.~G.~Meissner,
  %``Hyperon nucleon interactions: A chiral effective field theory approach,''
  Nucl.\ Phys.\  A {\bf 779} (2006) 244
  [arXiv:nucl-th/0605050];   J.~Haidenbauer, U.~G.~Meissner, A.~Nogga and H.~Polinder,
  %``The hyperon nucleon interaction: Conventional versus effective field theory
  %approach,''
  arXiv:nucl-th/0702015.



\bibitem{qnp}
  C.~Hanhart,
  %``Towards an understanding of the light scalar mesons,''
  arXiv:hep-ph/0609136.


\bibitem{disptheo}
N.~I.~Muskhelishvili,
{\it Singular Integral Equations},
(P.~Noordhof N.~V., Groningen, 1953);
R.~Omnes,
%``On The Solution Of Certain Singular Integral Equations Of Quantum Field Theory,''
Nuovo Cim.\  {\bf 8}, 316 (1958);
W.~R.~Frazer and J.~R.~Fulco,
%``Effect Of A Pion-Pion Scattering Resonance On Nucleon Structure,''
Phys.\ Rev.\ Lett.\  {\bf 2}, 365 (1959).

\bibitem{mitachot1}
  A.~Gasparyan, J.~Haidenbauer, C.~Hanhart and J.~Speth,
  %``How to extract the Lambda N scattering length from production  reactions,''
  Phys.\ Rev.\  C {\bf 69}, 034006 (2004)
  [arXiv:hep-ph/0311116].

\bibitem{mitachot2}
  A.~Gasparyan, J.~Haidenbauer and C.~Hanhart,
  %``Extraction of scattering lengths from final-state interactions,''
  Phys.\ Rev.\  C {\bf 72}, 034006 (2005)
  [arXiv:nucl-th/0506067].

\bibitem{mitachot3}
  A.~Gasparyan, J.~Haidenbauer, C.~Hanhart and K.~Miyagawa,
  %``Lambda N scattering length from the reaction gamma d --> K+ Lambda n,''
  arXiv:nucl-th/0701090.


\bibitem{sechi}
F.~Eisele et~al.
Phys. Lett. B {\bf 37}, 1971 (204).


\bibitem{saclay} R. Siebert et al., 
Nucl. Phys. {\bf A567}, 819 (1994). 


\bibitem{tofprop}
A. Gillizer et al., in preparation.

\bibitem{jlab}
B.L. Berman et al., CEBAF proposal PR-89-045 (1989).

\bibitem{mami}
R.~Beck and A.~Starostin, 
Eur. Phys. J. A {\bf S19}, 279 (2004);
  R.~Beck,
  %``New results and future plans with real photons at MAMI,''
  Prog.\ Part.\ Nucl.\ Phys.\  {\bf 55}, 91 (2005).

\bibitem{elsa}
V. Kleber and H. Schmieden, private communication. 

\end{thebibliography}
 \end{document}